\magnification=1200
\overfullrule=0pt
\nopagenumbers

\baselineskip=14pt
\vsize 25 truecm
\hsize 17 truecm
\hoffset -0.5 truecm
\voffset -0.3 truecm
\input epsf

\vbox to 4 truecm {}
\centerline{\bf PROBING THE PHOTON STRUCTURE FUNCTION} \par
\centerline{\bf IN PHOTOPRODUCTION} \par
\medskip
\centerline{\bf M. Fontannaz} \par \smallskip
\centerline{Laboratoire de Physique Th\'eorique et Hautes
Energies\footnote{*}{Laboratoire associ\'e au
Centre National de la Recherche Scientifique - URA 63}} \par
\centerline{Universit\'e de Paris XI, b\^atiment 211,
91405 Orsay Cedex, France} \vbox to 2 truecm {}
\vskip 5 mm
\noindent Abstract : \par
We show how the parton distributions in the photon can be accurately
measured in the photoproduction of large-$p_{\bot}$ jets at HERA. A
short review is given of the beyond Leading Logarithm formalism for the
photon structure function, with a discussion of the non perturbative
input. \par
\vbox to 4 truecm {}
\noindent LPTHE Orsay 94/65 \par
\vbox to 2 truecm {}
\centerline{\it Invited Talk given at the XXIXth Rencontres de Moriond, March
1994}
\vfill \supereject
\noindent {\bf 1. Introduction}
\vskip 3 mm
\baselineskip=18pt
Hard processes in photoproduction at HERA are ideal reactions to observe the
parton distributions
in the real photon. Figure 1 schematically shows such a reaction in which a
photon coming from the
incident electron interacts with the photon and produces large-$p_{\bot}$ final
hadrons. An
interesting point of the photoproduction reactions at HERA is the fact that the
photon is almost
real, because of the tagging condition on the electron$^{1), 2)}$.
$$
\epsfbox{des1}
$$
Actually the photon interacts with the partons of the proton either directly
(Fig. 2a), or via its
partonic component (Fig. 2b). This latter contribution, often called resolved,
is dominant at small
transverse momentum $p_{\bot}$, as we shall see below, and progressively
decreases in comparison
with the direct contribution which wins at large $p_{\bot}$. The CM energy
available at HERA allows
to explore a completely new kinematical domain, distinct from the one already
observed in fixed
target experiments$^{3), 4), 5), 6)}$ in which the direct contribution is
important.
$$
\epsfbox{des2}
$$
Until now our knowledge of the quark distribution in photon comes from the deep
inelastic
scattering of a virtual photon ($Q^2 = - q^2 >> \Lambda^2$) on a real photon
($p^2 = 0$). Numerous
data have been obtained at PEP, PETRA, TRISTAN and LEP$^{7)}$. They are however
not very accurate~;
although constraining the quark distributions, they hardly allow a quantitative
comparison with
theory. \par

 On the other hand, future results, not only at HERA, but also at TRISTAN in
photon-photon
collisions should allow a quantitative determination of the gluon and quark
distributions in the
photon, and this fact has triggered an important theoretical activity. Beyond
leading logarithm
parton distributions in the photon have been calculated, as well as higher
order QCD corrections to
various hard subprocesses. \par
In this talk, I will discuss these two points. Let us consider the reaction of
Fig. 2b which can be
symbolically written

$${d \sigma^{jet} \over d \vec{p}_{\bot} dy} = \sum_{ij} {\cal P}_{\gamma}^i
\otimes \widehat{\sigma}
\left ( ij \to jet \ X \right ) \otimes G_P^j \eqno(1)$$

\noindent where ${\cal P}_{\gamma}^i$ and $G_P^j$ are the parton distributions
in the photon and
in the proton, and where $\widehat{\sigma}$ is the subprocess cross-section.
Expanding ${\cal
P}_{\gamma}^i$ and $\widehat{\sigma}$ in power of $\alpha_s$, we obtain an
expression
$${d \sigma^{jet} \over d \vec{p}_{\bot} \ d \eta} = \sum_{i,j} \left ( {4 \pi
\over
\alpha_s(p_{\bot}^2)} a_i + b_i \right ) \otimes \left ( \alpha \
\alpha_s(p_{\bot}^2)
\widehat{\sigma}_{ij}^{BORN} + \alpha \ \alpha_s^2(p_{\bot}^2) K_{ij} \right )
\otimes
G_P^j \eqno(2)$$

\noindent which shows the Leading Logarithm (LL) contributions to the jet
cross-section (associated
with $a_i$ and $\widehat{\sigma}_i^{BORN}$ which describe the $2 \to 2$
subprocesses), and the
Higher Order (HO) QCD corrections coming from $b_i$ and $K_{ij}$. I do not
discuss the well-known
parton distributions in the proton and concentrate on the incident photon. \par

The term $b_i$ describes the effects of the HO corrections to the evolution
equations of the quark
and gluon distributions in the photon ; it is discussed in the next section in
which we also
address the issue of the non perturbative part of these distributions. \par

In section 3 I will discuss the HO corrections $K_{ij}$ corresponding to $2 \to
3$ subprocesses and
virtual corrections to $2 \to 2$ subprocesses, and show how they make the
inclusive jet cross-section
more stable with respect to variations of the factorization scale. This fact
makes possible a
quantitative comparison between theoretical predictions and data. Other final
states have also
been studied and the effects of the HO corrections calculated~: large
$p_{\bot}$ final photons$^{8),
9)}$ and hadrons$^{10), 11), 12), 13)}$, heavy flavors and massive lepton
pairs$^{14)}$. Similar
calculations in $\gamma \gamma$ collisions exist for the production of jets,
hadrons and heavy
flavors$^{15)}$. I will here concentrate on the jet production at HERA$^{16),
17), 18), 19), 20),
21)}$ and TRISTAN$^{22)}$, because these reactions can be compared with already
existing data.

 \vskip 3 mm \noindent {\bf 2. The photon structure function} \vskip 3 mm
The parton contents of the photon$^{7)}$ can be measured in deep inelastic
scat\-te\-ring experiments
in which the virtual photon $\gamma^{\ast}$ of momentum $q$ ($Q^2 = - q^2 >>
\Lambda^2$) probes the
short distance behavior of the real photon $\gamma$ of momentum $p$. The
structure function
$F_2^{\gamma}$ of this reaction is proportional, in the LL approximation, to
the quark distributions
in the real photon
$$F_2^{\gamma}(x, Q^2) = x \sum_{f=1}^{n_f} e_f^2 \left ( q_{\gamma}^f(x, Q^2)
+
\bar{q}_{\gamma}^f(x, Q^2) \right ) \ \ \ . \eqno(3)$$

\noindent  The sum in (3) run over the quark flavors and $x = Q^2/2p.q$. \par

It is instructive to consider the contribution to $F_2^{\gamma}$ of the lowest
order diagrams of
Fig.~3. Contrarily to the case of a hadronic target, the lower part of the
diagram is known~: it is
given by the coupling of photon to quark.
$$
\epsfbox{des3}
$$
\noindent This contribution is therefore exactly calculable, with the following
result
for a quark of charge $e_f$~:
$$F_2^{\gamma}(x, Q^2) = 3e_f^4 {\alpha \over \pi} x \left \{ \left ( x^2 +
(1-x)^2 \right )
\ell n {Q^2 \over m_f^2} + \left ( x^2 + (1-x)^2 \right ) \ell n {1-x \over x}
+ 8x(1-x) -1
\right \} . \eqno(4)$$

However our result (4) is not directly related to a physical process, because
it depends on the
unknown quark mass $m_f$, used as a cut-off to regularize a logarithmic
divergence. Actually this
perturbative approach is certainly not valid when the virtuality $|k|^2$ of the
exchanged quark
becomes small. We then go into a non perturbative domain where we lack
theoretical tools and we must
resort to models to describe non perturbative (NP) contributions to
$F_2^{\gamma}$. A popular model
is the ``Vector Meson Do\-mi\-nan\-ce Model'' (VDM) which consider that the
real photon couple to
vector mesons. Therefore the real photon, besides a direct coupling to a $q
\bar{q}$ pair,
has a VDM component which is also probed by the virtual photon. \par

The latter component contributes to $F_2^{\gamma}$ and must be added to
expression (4). Keeping only
the term in (4) proportional to $Log \ Q^2/m_f^2$ (LL approximation), we write
$$F_2^{\gamma}(x, Q^2) = 3 e_f^4 {\alpha \over \pi} x \left ( x^2 + (1 - x)^2
\right ) \ell n {Q^2
\over Q_0^2} + x \sum_{V=\rho, \omega , \phi} e_f^2 \left ( q_f^V(x) +
\bar{q}_f^V(x) \right ) \ \ \
. \eqno(5)$$

\noindent The scale $Q_0^2$ is the value of $Q^2$ at which the perturbative
approach is no more
valid. The perturbative contribution vanishes and $F_2^{\gamma}$ is described
only by the non
perturbative contribution $q_f^{NP}(x) = q_f^V(x)$ which describes the quark
contents of vector
mesons. \par

We have to keep in mind that this way of treating the non perturbative part of
$F_2^{\gamma}$ is
due to our lack of theoretical understanding of this contribution. There are
other
approaches$^{23)}$, especially that of ref. 24) which takes into account the
interaction between the
quarks and the gluon condensate. These different approaches must ultimately be
compared with
experiment. \par

QCD corrections to the diagrams of Fig. 3 do not change the basic structure of
expression (5). In
the LL approximation$^{25)}$, the perturbative quark distribution is given by
the sum of ladder
diagrams (Fig. 4) (for simplicity we forget the gluons and consider only the
non singlet quark
distribution)
$$
\epsfbox{des4}
$$
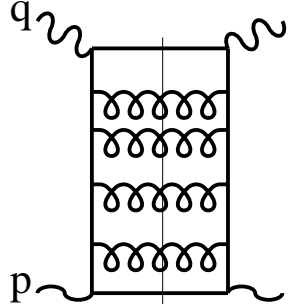
\centerline{Fig. 4 : Ladder diagram contribution to $F_2^{\gamma}$ (the thin
line cuts final
partons).}
\vskip 3 mm

\noindent  and the non perturbative part acquires a $Q^2$-dependence which is
identical to that
of a quark distribution in a hadron. The sum of the ladder diagrams can be
written in a very compact
form (AN is for anomalous, a designation introduced in ref. 25)

$$q_{\gamma}^{AN}(n, Q^2) = {\alpha \over 2 \pi}
\int_{\alpha_s(Q_0^2)}^{\alpha_s(Q^2)} {d \alpha '_s
\over \beta (\alpha '_s)} k^{(0)}(n) e^{\int_{\alpha '_s}^{\alpha_s(Q^2)} {d
\alpha ''_s \over
\beta(\alpha ''_s)} {\alpha ''_s \over 2 \pi} P^{(0)}(n)} \eqno(6)$$

\noindent in terms of moments of the Altarelli-Parisi kernels $k^{(0)}(n) =
\int_0^1 dx \ x^{n-1}
k^{(0)}(x)$ and $P^{(0)}(n)$ describing the splitting of a photon into a $q
\bar{q}$ pair (the
bottom rung of the ladder) and the splitting of a quark into a quark and a
gluon (the other
rungs of the ladder). The beta function has the usual definition $\partial
\alpha_s(Q^2)/\partial \ell n \ Q^2 = \beta (\alpha_s) = - \alpha_s ({\alpha_s
\over 4 \pi} \beta_0
+ ({\alpha_s \over 4 \pi})^2 \beta_1 + \cdots )$. The total quark distribution
is given by
$$q_{\gamma}(n, Q^2) = q_{\gamma}^{AN}(n, Q^2) + q_{\gamma}^{NP}(n, Q^2)
\eqno(7)$$

\noindent  which verifies the inhomogeneous equation

$$Q^2 {\partial q_{\gamma}(n, Q^2) \over \partial Q^2} = {\alpha \over 2 \pi}
k^{(0)}(n) +
{\alpha_s(Q^2) \over 2 \pi} P^{(0)}(n) q_{\gamma}(n, Q^2) \ \ \ . \eqno(8)$$

\noindent  As in (5) we introduce the boundary condition$^{27)}$ $Q_0^2$ in (6)
so that
$q_{\gamma}^{AN}(n, Q_0^2)$ vanishes when $Q^2 = Q_0^2$. \par

The modifications of these LL results due to HO QCD corrections$^{27)-30)}$ are
obtained by replacing the LL kernels of (6) by kernels involving HO
contributions
$$\eqalignno{
k(n) & = {\alpha \over 2 \pi} k^{(0)}(n) + {\alpha \over 2 \pi} {\alpha_s \over
2 \pi} k^{(1)}(n)
+ \cdots \cr
P(n) & = {\alpha_s \over 2 \pi} P^{(0)}(n) + \left ( {\alpha_s \over 2 \pi}
\right )^2 P^{(1)}(n)
+ \cdots \ \ \ , &(9) \cr
}$$
\noindent  and by a modification of the expression of $F_2^{\gamma}$ in terms
of parton
distributions (the gluon contribution is now explicitly written)
$$F_2^{\gamma}(n - 1, Q^2) = e_f^2 C_q(n, Q^2) \left ( q_{\gamma}(n, Q^2) +
\bar{q}_{\gamma}(n,
Q^2) \right ) + C_g(n, Q^2) g_{\gamma}(n, Q^2) + C_{\gamma}(n) \eqno(10)$$
\noindent  where $C_{\gamma}$ is the ``direct term'', given by the part of (4)
not proportional to
$\ell n {Q^2 \over m_f^2}$. $C_q$ and $C_g$ are the well-known Wilson
coefficients which are
identical to those found in the case of a hadronic target. \par

Several parametrizations of the quark and gluon distributions in the photon are
now
available$^{31), 32), 33)}$, which take into account HO QCD corrections to the
Altarelli-Parisi
kernels (9). A comparison between data$^{34)}$ and theoretical
predictions$^{33)}$ is shown in
Fig. 5. We see that we get a reasonable agreement between theory and
experiment. \par
 \vbox to 7 cm {}
 \baselineskip = 14 pt \noindent Fig. 5 : Comparison of JADE data with
theoretical predictions. The
dashed curve cor\-res\-ponds to a non perturbative (VDM) input set equal to
zero.
\vskip 3 mm

\baselineskip=18 pt
More comparisons can be found in ref. 31) to 33). From this study we conclude
that the quark
distribution in the photon is constrained by data on $F_2^{\gamma}$, but that
they are not accurate
enough to teach us something about the non perturbative contribution and the
value of $Q_0^2$. \par

A delicate point when working beyond the LL approximation is that of the
factorization scheme. A
change in the factorization scheme is translated into a change in $k^{(1)}$ and
$C_{\gamma}$
but in such a way that the physical quantity $F_2^{\gamma}$ remains unmodified
(at order
$\alpha_s^0$). On the other hand $q^{\gamma}$ is not an invariant with respect
to the factorization
scheme and a change in $k^{(1)}$ causes modifications in $q_{\gamma}^{AN}$
\underbar{and}
$q_{\gamma}^{NP}$. Therefore the separation (7) in a perturbative and a
 non perturbative part is
\underbar{not} scheme invariant and the statement that $q_{\gamma}^{NP}$ can be
described by VDM has
no meaning, unless one specifies in which factorization scheme it is valid.
This problem is
discussed in details in ref. 33).

\vskip 3 mm \noindent {\bf 3. Jets in photoproduction} \vskip 3 mm
A quantitative analysis of data cannot be performed without the knowledge of
theoretical
cross-sections including HO QCD contributions. The reason is the scale
dependence of the
cross-section, shown in Fig. 6a, for a calculation done in the LL
approximation. The scale $\mu$
is the renormalization scale, argument of $\alpha_s(\mu)$, and $M$ is the
factorization scale
appearing in the distribution functions. The LL jet cross-section is a
decreasing function of the
scales and we do not know for which values of $\mu$ and $M$, we have to compare
predictions with
data. \par

The beyond LL cross-section is much more stable with respect to variation of
$\mu$ and $M$, as it
can be seen in Fig. 6b. It is however not fully flat, as it should be if all
orders in $\alpha_s$
were included, and we have to keep this fact in mind when comparing with data.
\par

\vbox to 6.3 truecm {}
\hskip 4 truecm (a)\hskip 6 truecm (b)\par
\baselineskip = 14 pt \noindent  Fig. 6 : The jet cross-section $d
\sigma^{jet}/d \vec{p} d \eta$ at
$p_{\bot} = 10 \ {\rm GeV/c}$ and $\eta = 0$ as a function of $\ell n(M)$ and
$\ell n(\mu)$. a) LL
prediction, b) beyond LL prediction (rotated by $90^{\circ}$).
\vskip 3 mm
\baselineskip=18 pt
The almost cancellation of the scale dependence between the LL and beyond LL
parts of the
cross-section is easily understandable from the following example represented
in Fig.~7. The HO
corrections to the Compton subprocess of Fig. 7a, due to the emission of a
second gluon (Fig.
7b), is obtained by integrating over the momenta $p$ and $k_2$ (only the jet of
momentum $k_1$ is
observed).
$$
\epsfbox{des7}
$$
\noindent In the course of this calculation a collinear divergence appears
(corresponding to the
kinematical configuration of Fig. 7c) that we regularize by giving a mass to
the quark

$$K^{di \sigma} \simeq {\alpha \over 2 \pi} Log {p_{\bot}^2 \over m_q^2} \left
( x^2 + (1 - x)^2
\right ) \otimes \widehat{\sigma}^{BORN}(q \bar{q} \to gg) + \kappa (x) \ \ \ .
\eqno(11)$$

\noindent A part of this contribution is however already contained in the
resolved contribution of
Fig.~2b (calculated with the scale $M$). We have therefore to subtract this
contribution from
$K_{\gamma q}^{div}$ to get a finite $M$-dependent HO correction

$$K = K^{div} - {\alpha \over 2 \pi} Log {M^2 \over m_q^2} \left ( x^2 + (1 -
x)^2 \right )
\otimes \widehat{\sigma}^{BORN} (q \bar{q} \to gg)$$
$$= {\alpha \over 2 \pi} Log {p_{\bot}^2 \over M^2} \left ( x^2 + (1 - x)^2
\right ) \otimes
\widehat{\sigma}^{BORN} (q \bar{q} \to gg) + K(x) \eqno(12)$$

\noindent to the direct term of Fig. 7a. \par

{}From this calculation, we draw the following conclusions~: \par
1) there is a compensation between the $M$-dependence of the HO corrections to
the direct term and
that of the resolved contribution (at the order in $\alpha_s$ at which the HO
calculation is
done), \par
2) it is no more meaningful to distinguish between resolved and direct terms in
a BLL
calculation. For instance $K$ contains terms corresponding to resolved
kinematical
configurations, \par
3) each separate contribution to the jet cross-section is scale dependent. Only
the sum has a
physical meaning, and is approximately scale independent. \par

The starting of HERA has triggered several BLL calculations of the jet
cross-\break
\noindent sections$^{16), 17), 18), 19), 20), 21)}$ that can be compared with
already existing data.
Let us here compare the predictions of ref. 21), obtained the choice of scale
$\mu = M = p_{\bot}$,
with H1 data$^{35)}$. In Fig. 8 and 9, we see the jet cross-section as a
function of the
pseudo-rapidity $\eta$, integrated over $p_{\bot} > 7 \ {\rm GeV/c}$. The H1
data have large error
bars and a $40 \ \%$ normalization uncertainty is not included, but it is not
obvious
to reconcile them with theory.

\vbox to 6.5 truecm {}
{\baselineskip = 14 pt \noindent  Fig. 8 : Jet cross-section as a function
\hskip 1.5 truecm Fig. 9 :
Same as fig. 8. Gluon distribution \hfill
of $\eta$ : full line. Direct contribution : dots, \hskip 1.1 truecm in the
photon = 0 :
dashed line, gluon in \hfill
\noindent VDM input = 0 : dashed line. \hskip 3.1 truecm the proton = 0 : dots.
\par}
 \vskip 3 mm

\baselineskip = 18 pt \noindent  We could try to do it by modifying the gluon
contents of the proton
and photon (the sensitivity of $d \sigma^{jet}/d \eta$ to these gluon contents
is shown in Fig. 9),
but it is too early in view of the quality of the data. We can however already
notice that accurate
experimental results will put severe constraints on the gluon distributions in
the photon and
proton, and on the non-perturbative input of the parton distributions in the
photon (Fig. 8).
\par
\vskip 3 mm
\noindent {\bf 4. Jets in photon-photon collision}
\vskip 3 mm
 The jet production is the collision of two real photons is also very sensitive
to the parton
distributions in the photon, which can intervene twice. HO corrections to the
jet cross-section
$d \sigma (\gamma \gamma \to jet \ X)/d \vec{p} \ d \eta$ have been recently
calculated, and the
effects of the convolution with the Weizs\"acker-Williams photon spectrum in
the electron$^{36)}$
carefully discussed$^{22)}$. The resulting jet cross-section can be compared
with recent TOPAZ
data$^{37)}$ from TRISTAN and in Fig.~10 we can see the good agreement between
theory and
experiment. We also notice the large contribution coming from the partons in
the photon and the
facts that the cross-section is not much sensitive to the non-perturbative
input of the parton
distributions$^{22), 38)}$. Therefore the jet production in photon-photon
collisions constrains the
parton distributions in the photon, but does not allow a determination
of the non perturbative input.

\vbox to 7 truecm {}
 \baselineskip = 14 pt \noindent Fig. 10 : TOPAZ data on inclusive jet
production and theoretical
predictions for\break \noindent $\int_{.7}^{.7} d \eta {d \sigma^{e^+e^--jet}
\over dp_T \ d \eta}$.
The top curve is the theoretical prediction based on the standard photon
structure functions, the
middle one is based on structure functions with half the VDM input, and the
lower one is based on the
perturbative component only. The dash-dotted curve is the ``direct
contribution''.

 \baselineskip = 14 pt

\vskip 3 mm
\noindent {\bf References}
\vskip 3 mm
\item {1.}  T. Ahmed et al., H1 Collab., Phys. Lett. \underbar{B299} (1993)
374.
\item{2.} M. Derrick et al., ZEUS Collab., Phys. Lett. \underbar{B322} (1994)
287.
\item {3.} R. Barate et al., Phys. Lett. \underbar{B182} (1986) 409,
\underbar{B174} (1986) 458,
\underbar{B168} (1986) 163.
\item{4.} P. Astbury et al., Phys. Lett. \underbar{B152} (1985) 419.
\item {5.} R. J. Apsimon et al., Z. Phys. \underbar{C43} (1989) 63.
\item {6.} J. R. Raab et al., Phys. Rev. \underbar{D37} (1988) 2391.
\item{} J. C. Anjos et al., Phys. Rev. Lett. \underbar{62} (1989) 513.
\item {7.} Reviews on the photon structure function may be found in
\item{} C. Berger and W. Wagner, Phys. Rep. \underbar{146} (1987) 1 ;
\item{} H. Kolanski and P. Zerwas, in High Energy $e^+e^-$ physics, World
Scientific, Singapore,
1988, eds. A. Ali and P. S\"oding ;
\item{} J. H. Da Luz Vieira and J. K. Storrow, Z. Phys. \underbar{C51} (1991)
241.
\item {8.} P. Aurenche, P. Chiappetta, M. Fontannaz, J. Ph. Guillet and E.
Pilon, Z. Phys.
 \underbar{C56} (1992) 589.
\item {9.} L. E. Gordon and J. K. Storrow, preprint DO-TH 93/13.
\item {10.} F. M. Borzumati, B. A. Kniehl and G. Kramer, Z. Phys.
\underbar{C59} (1993) 341 ;
\item{} B. A. Kniehl and G. Kramer, preprint DESY 94-009.
\item{11.} L. E. Gordon, preprint DO-TH 93/25.
\item {12.} M. Greco, S. Rolli and A. Vicini, FERMILAB-Pub-94/075-A.
\item {13.} P. Aurenche, M. Fontannaz, M. Greco and J. Ph. Guillet, in
preparation.
\item {14.} M. Gl\"uck, E. Reya and A. Vogt, Phys. Lett. \underbar{B285} (1992)
285.
\item{15.} M. Dress, M. Kr\"amer, J. Zunft and P. M. Zerwas, Phys. Lett.
\underbar{B306} (1993) 371.
\item{16.} L. E. Gordon and J. K. Storrow, Phys. Lett. \underbar{B291} (1992)
320.
\item{17.} M. Greco and A. Vicini, Nucl. Phys. \underbar{B405} (1994) 386.
\item{18.} D. B\"odeker, Z. Phys. \underbar{C59} (1993) 501, Phys. Lett.
\underbar{B292} (1992) 164.
\item{19.} G. Kramer and S. G. Salesch, Z. Phys. \underbar{C61} (1994) 277.
\item{20.} D. B\"odeker, G. Kramer and S. G. Salesch, DESY 94-042.
\item{21.} P. Aurenche, M. Fontannaz, J.-Ph. Guillet, in preparation.
\item{22.} P. Aurenche, J.-Ph. Guillet, M. Fontannaz, Y. Shimizu, J. Fujimoto,
K. Kato, KEK
preprint 93-180, LAPP-TH 436/93, LPTHE-Orsay 93/47.
 \item{23.} I. Antoniadis and G. Grunberg, Nucl. Phys. \underbar{B213} (1983)
455.
\item{24.} A. S. Gorski, B. L. Ioffe, A. Yu Khodjaminian, A. Oganesian, Z.
Phys. \underbar{C44}
(1989) 523.
 \item{25.} E. Witten, Nucl. Phys.  \underbar{B210} (1977) 189.
\item{26.} A review on QCD corrections, and previous references on the subject
can be found in~: F.
M. Borzumati and G. A. Schuler, Z. Phys. \underbar{C58} (1993) 139.
\item{27.} M. Gl\"uck and E. Reya, Phys. Rev. \underbar{D28} (1983) 2749.
\item{28.} W. A. Bardeen and A. J. Buras, Phys. Rev. \underbar{D20} (1979) 166
; \underbar{21}
(1980) 2041 (E).
 \item{29.} M. Fontannaz and E. Pilon, Phys. Rev. \underbar{D45} (1992) 382.
\item{30.} E. Laenen, S. Riemersma, J. Smith and W. L. van Neerven, University
of Leiden preprint
ITP-SB-93-46.
\item{31.} L. E. Gordon and J. K. Storrow, Z. Phys. \underbar{C56} (1992) 307.
\item{32.} M. Gl\"uck, E. Reya and A. Vogt, Phys. Rev. \underbar{D45} (1992)
3986, \underbar{D46}
(1992) 1973.
\item{33.} P. Aurenche, M. Fontannaz and J.-Ph. Guillet, preprint
ENSLAPP-A-435-93, LPTHE Orsay
93-37. \underbar{}
\item{34.} JADE Collab., W. Bartel et al., Z. Phys. \underbar{C24} (1984) 231.
\item{35.} I. Abt et al., H1 Collab., Phys. Lett. \underbar{B314} (1993) 436.
\item{36.} C. F. Weizs\"acker, Z. Phys. \underbar{88} (1934) 612.
\item{} E. J. Williams, Phys. Rev. \underbar{45} (1934) 729.
\item{37.} H. Hayashi et al., TOPAZ Collab., Phys. Lett. \underbar{B314} (1993)
141.
\item{38.} M. Fontannaz, Invited talk at the Meeting ``Two-Photon Physics from
DAPHNE to LEP 200
and Beyond'', Paris, Feb. 1994, preprint LPTHE Orsay 94/44.

\bye